\documentclass[prl,twocolumn,showpacs,superscriptaddress,floatfix]{revtex4}
\usepackage{graphics,epsfig,amsfonts,amssymb,amsmath,ulem,color}
\begin{document}
\title{Correlation, doping and interband effects on the optical conductivity of iron superconductors}
\author{M.J. Calder\'on}
\affiliation{Instituto de Ciencia de Materiales de Madrid, ICMM-CSIC, Cantoblanco, E-28049 Madrid (Spain).}
\author{L. de' Medici}
\affiliation{European Synchrotron Radiation Facility, BP 220, F-38043 Grenoble Cedex 9 (France)}
\affiliation{Laboratoire de Physique et Etude des Materiaux, UMR8213 CNRS/ESPCI/UPMC, Paris (France)}
\author{B. Valenzuela}
\affiliation{Instituto de Ciencia de Materiales de Madrid, ICMM-CSIC, Cantoblanco, E-28049 Madrid (Spain).}
\author{E. Bascones$^*$}
\affiliation{Instituto de Ciencia de Materiales de Madrid, ICMM-CSIC, Cantoblanco, E-28049 Madrid (Spain).}
\email{leni@icmm.csic.es}
\date{\today}
\begin{abstract}
Electronic interactions in multiorbital systems lead to non-trivial features in the optical spectrum. In iron superconductors the Drude weight is strongly suppressed with hole-doping. We  discuss why the common association of the renormalization of the Drude weight with that of the kinetic energy, used in single band systems,  does not hold in multi-orbital systems.  This applies even in a Fermi liquid description when each orbital is renormalized differently, as it happens in iron superconductors. We estimate the contribution of interband transitions at low energies. We show that this contribution is strongly enhanced by interactions and dominates the coherent part of the spectral weight in hole-doped samples at frequencies currently used to determine the Drude weight.

\end{abstract}
\maketitle

The importance of electronic correlations in iron superconductors has been debated since their discovery~\cite{kamihara2008}. 
Depending on whether the emphasis was on their metallicity 
or on their low conductivity these materials 
were initially described in terms of weakly 
correlated~\cite{raghu2008,mazin2008,chubukov2008,cvetkovic09} 
or almost localized electrons~\cite{yildirim2008,si2008}. 
Experimental evidence places iron superconductors in
between both descriptions~\cite{lu2008,qazilbash09,dagottonat12}. 
More recently, the relevance of the multi-orbital character in iron superconductors has been revealed by Dynamical Mean Field Theory~\cite{shorikov09,haule09,liebsch2010} and slave spins~\cite{si2012,demedici2014} calculations: the electronic correlations can be orbital dependent and are highly influenced by Hund's coupling. 
Undoped compounds accomodate $6$ electrons 
in the $5$ Fe-d orbitals. For this filling, $J_H$ favors a 
bad metallic state~\cite{werner2008,demedici11,liebsch2010b,si2012} with strongly renormalized bands, 
usually referred to as a Hund 
metal~\cite{haule09}.

The Hund metal picture predicts an asymmetry 
in correlations upon electron or hole doping~\cite{liebsch2010,liebsch2010b,werner2012,bascones2012,demedici2014,nosotrasprb12}. 
Hole-doping enhances correlations, i.e. the renormalization increases, 
as it approaches a Mott insulating phase at half-filling 
(5 electrons in 5 orbitals). 
With electron-doping the material moves away
from half-filling and correlations decrease. 
Hund's coupling also leads to orbital decoupling~\cite{demedici2009,demedici11-2}: In iron superconductors 
the d-orbitals 
have different fillings and bandwidths resulting in different 
renormalizations. This orbital differentiation has been shown to 
play an important role in the magnetic state~\cite{bascones2012}. 
Experimentally there is increasing evidence for orbital 
differentiation and increasing correlations with hole-doping~\cite{demedici2014} 
from specific heat~\cite{hardy2013-meingast}, quantum oscillations~\cite{terashima2013} and angle 
resolved photoemission~\cite{yoshida2012,sudayama2011,yi2013-shen,xu2013,maletz2014-borisenko} experiments, among others.

Optical conductivity is a useful tool to analyze the electronic properties of strongly correlated electron systems~\cite{basovreview}. Many of the  experiments focus on the suppression of the Drude  weight or on the scattering rate and mass renormalizations obtained from fittings of the low energy spectrum. The recipe for extracting data from the spectrum is well established for single band systems but for multi-band materials, as the iron superconductors, the different features are difficult to disentangle for two main reasons: First, the contribution of the different orbitals to the Drude weight has to be considered and, second, the interband transitions  in iron superconductors contribute at relatively low energies~\cite{vanHeumenEPL10,benfatto2011,nosotrasprb13,marsik2013,charnukha14}.

The optical spectrum of iron pnictides is characterized by a zero energy peak  followed by a plateau-like region and a bump.  There have been different attempts to fit and explain these features~\cite{yin11, charnukha14}. The bump is usually associated to interband transitions. The  interpretation of the low energy part of the spectrum up to frequencies $1500-3000$ cm$^ {-1}$ (peak and plateau) is more controversial. Following procedures used in single band systems, it has been integrated to estimate the renormalization of the kinetic energy~\cite{qazilbash09,degiorgi2011}.  This low energy spectrum has been also
fitted using (i) a generalized Drude model~\cite{qazilbash09}, (ii) two (narrow and wide) Drude peaks~\cite{barisic-dressel2010,uchida2010,degiorgi10,carbotte10,akrap10,vandermarel10,min2013} (leading to an unphysical mean-free path $\sim$0.8 \r{A} for the wide peak)~\cite{charnukha14}, and (iii) one or two Drude peaks and interband transitions~\cite{vanHeumenEPL10, marsik2013}. Recent work has helped clarify which interband transitions are optically active\cite{benfatto2011,nosotrasprb13}. However, to date there is no estimate of how large is the impact of these  transitions at low energies.

Here we analyze the optical conductivity $\sigma^\prime (\omega)$ of iron superconductors as a function 
of doping.  Interactions are introduced within slave spin mean field.  Our approach  only includes the coherent contribution to the optical conductivity while addressing the renormalization of the quasiparticles. It emphasizes the role of the quasiparticle interband transitions, shifted to low energies by the 
overall squeezing of the bandstructure due to electronic correlations. 
We find that interband transitions give a non-negligible contribution to the low-energy plateau found in the optical spectrum of undoped compounds and account for a large fraction of the spectral weight at the cutoff frequencies currently used to determine the Drude weight. This fraction is strongly enhanced in hole-doped samples as the larger effect of interactions strongly suppresses the Drude weight. 
We analyze the relationship between the Drude weight and the kinetic energy and their renormalizations. We show that with orbital differentiation, the renormalization of both the Drude weight and the kinetic energy are not equal, not even within a Fermi liquid picture. Consequently, contrary to usual assumptions, the kinetic energy renormalization cannot be estimated from the Drude weight renormalization.  

We consider the model Hamiltonian discussed in Ref.~\cite{nosotrasprl10}. It contains the intraorbital $U$,  the interorbital $U'$, the Hund's coupling $J_H$, and the pair-hopping interaction $J'$. We assume that the 
equalities $U'=U-2J_H$ and $J_H=J'$, valid for rotational symmetric systems, hold. For the non-interacting part we have analyzed two 5-orbital tight-binding models~\cite{nosotrasprb09,graser09}. Both models give similar results.  The results plotted in the figures correspond to the tight-binding model proposed in Ref.~\cite{graser09}. 

\begin{figure}
\leavevmode
\includegraphics[clip,width=0.4\textwidth]{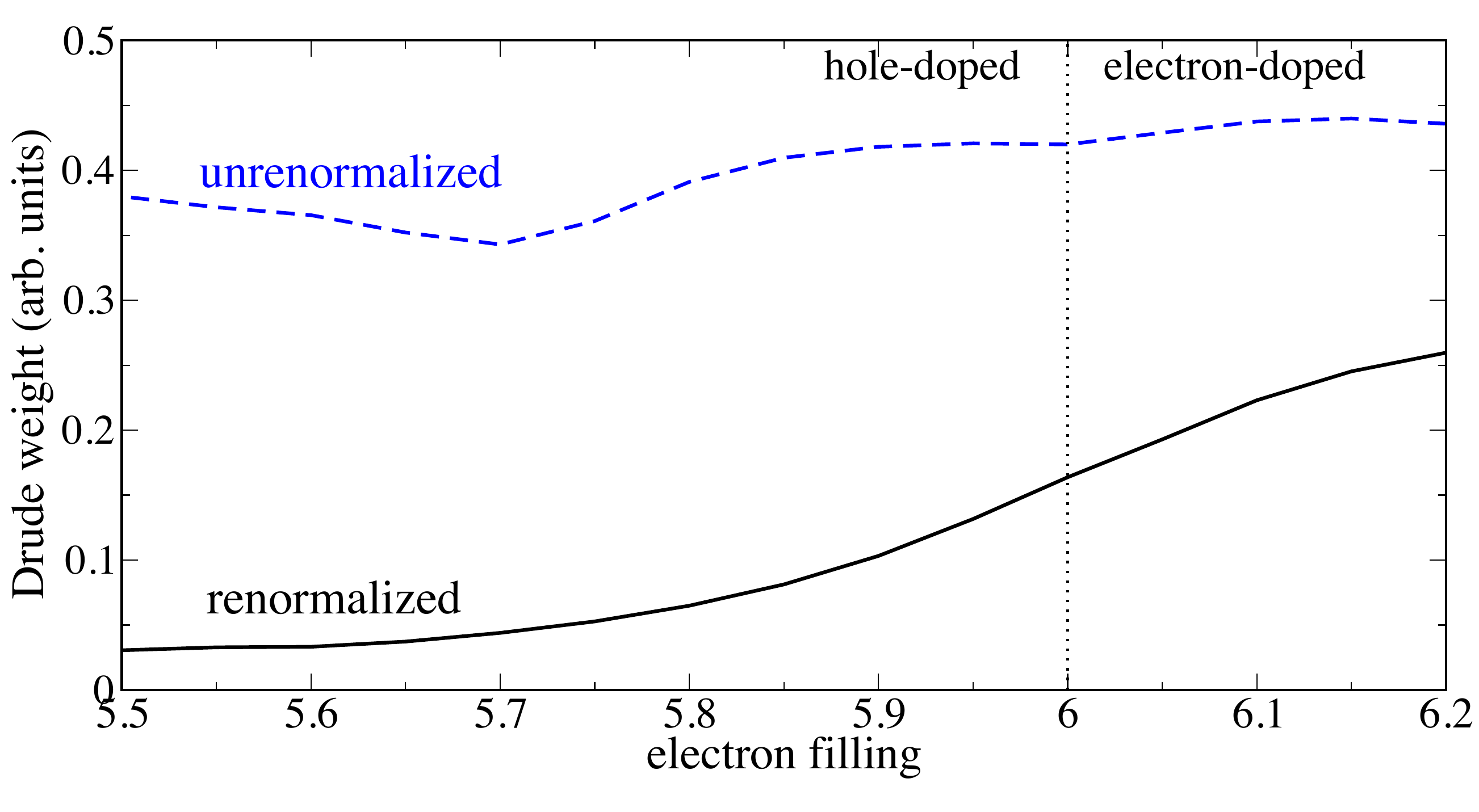}
\vskip -0.2cm
\includegraphics[clip,width=0.4\textwidth]{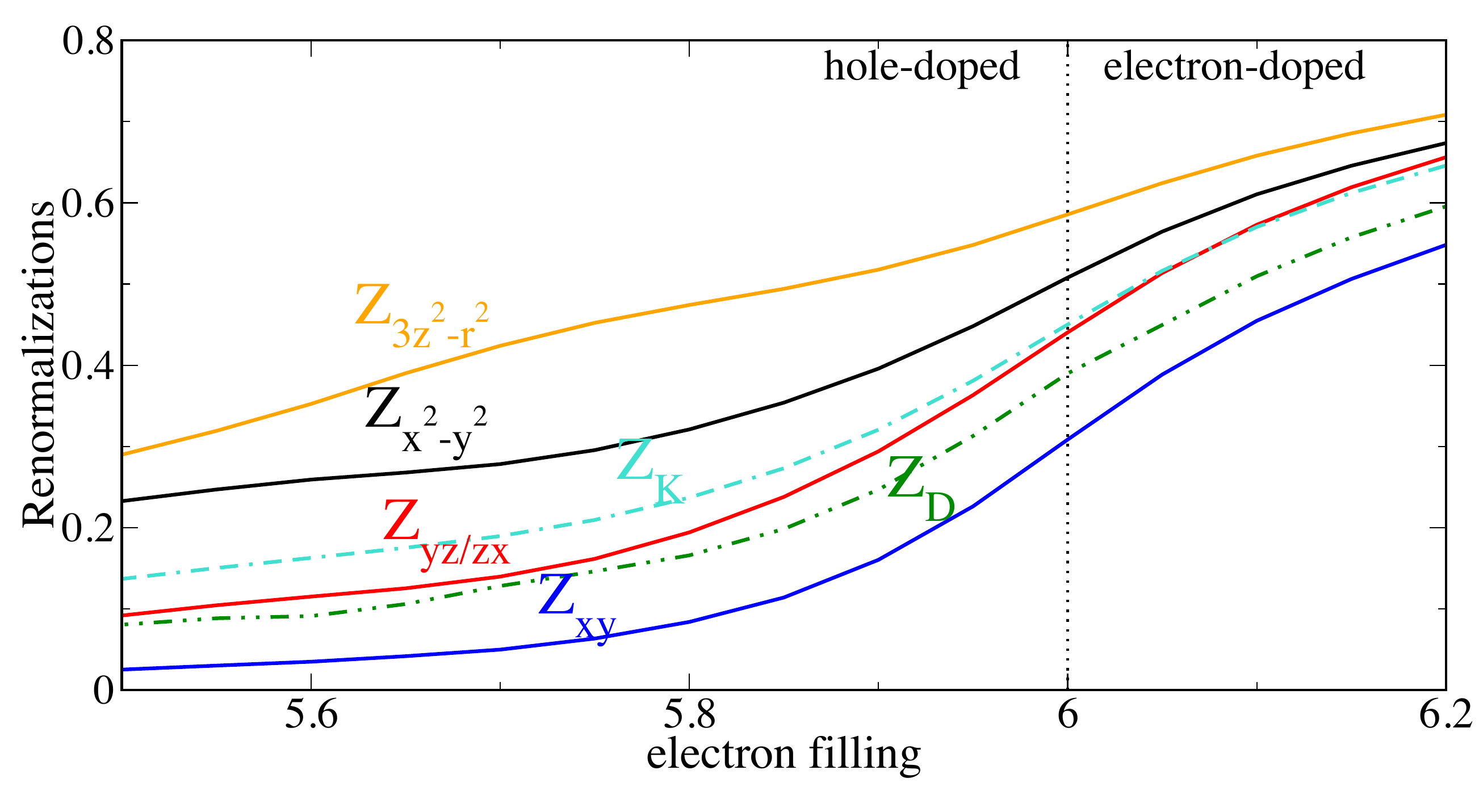}
\vskip -0.4cm
\caption{(Color online) (a) Drude weight corresponding to the non-interacting (top line) and interacting models (bottom). The Drude weight is strongly suppressed due to interactions, especially with increasing hole-doping. (b) Comparison among the doping-dependent orbital quasiparticle weights $Z_m$ (solid lines), the renormalized Drude weight $Z_D$, and the renormalized kinetic energy $Z_K$, see text for discussion.}
\label{fig:drude}
\end{figure}

Interactions are treated within the slave-spin mean field approximation~\cite{demedici05,demedici2010}. This method  accounts for electronic correlations through the orbital dependent quasiparticle weights $Z_m$,  see Fig.~\ref{fig:drude}(b), and shifts of the bare orbital energies. Only density-density interactions are considered, thus $J'$ does not enter in the calculations. We use $U=3$ eV and $J_H=0.25 U$, which give renormalization factors consistent with experiment for the undoped compounds for the considered bandstructure.  Doping is introduced via virtual crystal approximation, i.e. neither tight-binding nor interaction parameters of the original Hamiltonian are modified. All the calculations are performed at zero temperature. 

 The interactions are introduced via an effective renormalized Hamiltonian built with the $Z_m$  and onsite energies obtained from the slave-spin calculation, see Supplemental Material in Ref.~\cite{demedici2014} for details. In particular, each of the
hopping terms $t_{\bf i,j}^{\mu \nu}$ between orbitals $\mu$ and $\nu$ at sites ${\bf i}$ and ${\bf j}$ respectively are
 rescaled to $\sqrt{Z_\mu Z_\nu} \, t_{\bf i,j}^{\mu\nu}$.
 
\begin{figure}
\leavevmode
\includegraphics[clip,width=0.4\textwidth]{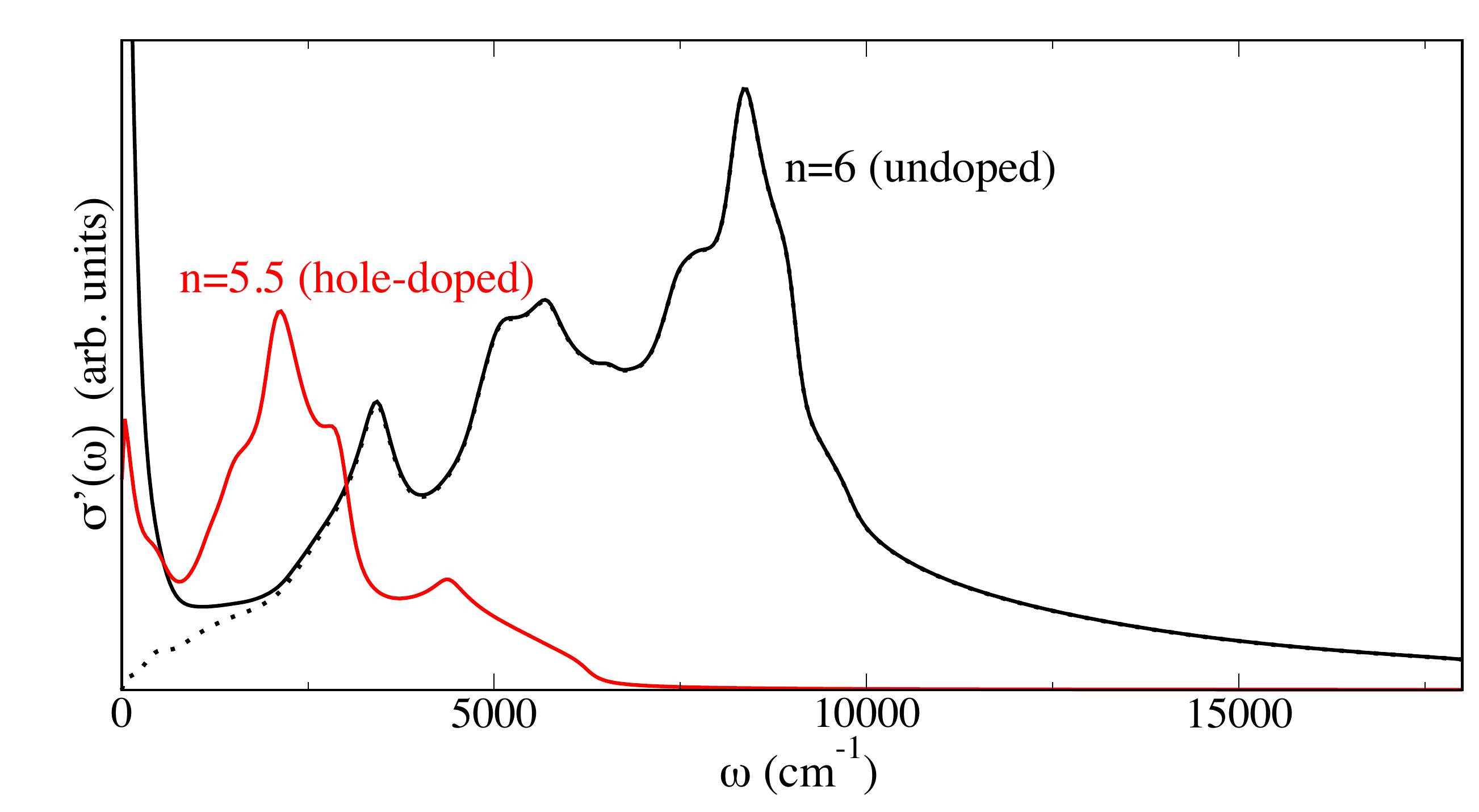}
\vskip -0.4cm
\caption{(Color online) Solid lines: Optical spectrum corresponding to 
an undoped ($n=6$) and a hole-doped system ($n=5.5$). Dotted line: Interband part of the optical spectrum for $n=6$. The $\delta$ functions which appear in the expressions 
for the optical conductivity have been broadened to lorentzians 
with half-width $\Gamma=160$ cm$^{-1}$ to mimic a constant scattering rate. Only 
the coherent contribution is included, see text.}
\label{fig:spectrum}
\end{figure} 

The optical conductivity and Drude weight are calculated for the renormalized Hamiltonian following the expressions derived in Ref.~\cite{nosotrasprb13} for multi-orbital systems. 
$\delta$ functions in these expressions are broadened into lorentzians with half-width $\Gamma$ to mimic the 
effect of the scattering rate which cannot be obtained in the slave spin calculation.  Only the coherent part 
of the electron enters. The incoherent part is expected to give a contribution 
weakly dependent on the frequency as Hund metals have wide Hubbard bands.

Fig.~\ref{fig:spectrum} shows the calculated spectrum for $n=6$ (undoped) and $n=5.5$ (hole-doped) systems.  As in experiments, three regions can be identified clearly in the 
spectrum of the $n=6$ system: A zero energy peak, a plateau-like region 
up to $~ 2000$ cm$^{-1}$, and a bump at higher energies. The bumps, 
which show a more complex structure than in experiments, would be smoothed by a frequency 
dependent scattering rate. 

Due to larger renormalization effects, hole 
doping shifts the interband transitions to smaller frequencies, the coherent 
contribution to the conductivity decreases and the shape of the spectrum is modified. 
The zero energy peak is more strongly suppressed than the high-energy part. The weakly frequency dependent 
incoherent contribution, not included here, would be more important in the hole doped system. A change in 
the spectrum with hole doping has also been observed experimentally~\cite{wang2012, uchida2013, uchida2013-2}.    

\begin{figure}
\leavevmode
\includegraphics[clip,width=0.4\textwidth]{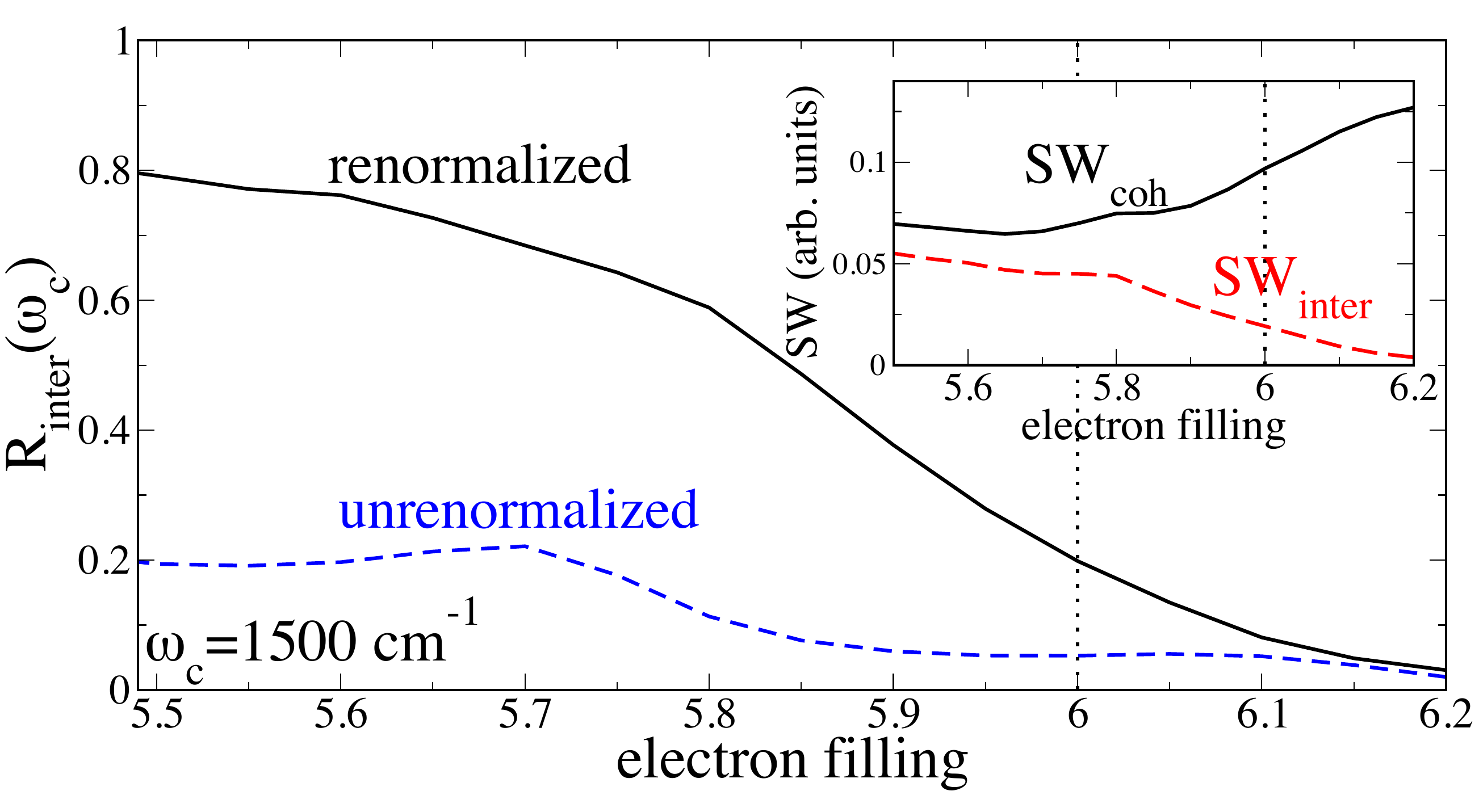}
\vskip -0.4cm
\caption{
(Color online) Main figure: Doping dependence of $R_{inter}(\omega_c)$, the ratio of the spectral weight coming from interband transitions to the total (coherent) spectral weight, for the non-renormalized and renormalized cases.  The optical conductivity is integrated up to a frequency cutoff $\omega_c=1500$ cm$^{-1}$. A broadening factor $\Gamma=160$ cm$^{-1}$ is used. The interband contribution amounts to one fifth of the total spectral weight in undoped compounds (6 electrons per Fe, vertical dotted line) and dominates the coherent part in hole-doped samples. The inset shows the total coherent spectral weight $SW_{coh}(\omega_c)$ and the interband part $SW_{inter}(\omega_c)$ for the renormalized case as a function of doping.}  
\label{fig:interbandsp}
\end{figure}

The zero energy peak and the high energy bump can be roughly ascribed to the Drude peak and the interband transitions respectively.
The low energy plateau has been interpreted differently in previous works: It has been fitted  to a wide Drude peak and adscribed to incoherence~\cite{wu-dressel2010,barisic-dressel2010,dai2013,uchida2013,uchida2013-2},  integrated together with the Drude peak~\cite{qazilbash09}, or partly related to interband transitions\cite{vanHeumenEPL10,benfatto2011,nosotrasprb13,marsik2013,charnukha14}.  The presence of this plateau in our calculations, which do not include the incoherent part, reveals that it cannot be completely associated with an incoherent continuum.  In fact,  interband transitions strongly contribute to the plateau, as shown by the dotted line in Fig.~\ref{fig:spectrum}. 

To analyze phenomenologically the contribution of interband transitions at low energies we integrate the spectral weight up to a frequency cutoff $\omega_c$  defining $ SW_{coh}(\omega_c)=SW_{Drude} (\omega_c)+SW_{inter} (\omega_c)$ and  separating the fraction $R_{inter}(\omega_c)=SW_{inter}(\omega_c)/SW_{coh}(\omega_c)$ coming from interband transitions.  $R_{inter}(\omega_c)$ is a quantity between zero and one: The closer to one, the larger the interband contribution to the spectral weight. Integration of the spectral weight up to a given cutoff is used to estimate the Drude weight from experiments. Fig.~\ref{fig:interbandsp} shows  $R_{inter}(\omega_c)$ as a function of doping for a frequency cutoff $\omega_c=1500$ cm$^{-1}$  often used experimentally. There is a strong increase  of $R_{inter}$ with hole-doping, which is more dramatic in the renormalized case. The overall 'squeezing' of the coherent bandstructure due to electronic correlations entails a reduction of all interband transition energies, thus shifting the corresponding spectral weight to lower frequencies. With a smaller frequency cutoff  $\omega_c=500$ cm$^{-1}$ (not shown) SW$_{inter}$ is still significant, evolving from $0.05$ at n=6 to $0.54$ at n=5.5.   Though the interband spectral weight $SW_{inter}(\omega_c)$ increases with hole-doping, the strong enhancement of $R_{inter}(\omega_c)$ is mostly due to the decrease of the total spectral weight $SW_{coh}(\omega_c)$, see inset in Fig.~\ref{fig:interbandsp}. This is due to the strong suppression of the Drude weight. 

The suppression of the Drude weight  due to the renormalization of the quasiparticle weight  is illustrated in  Fig.~\ref{fig:drude}(a) which compares the Drude weight as a function of doping for the non-interacting $D_{unren}$ and renormalized $D_{ren}$ tight-binding models.    $D_{unren}$ is non-monotonous and weakly dependent on doping,  presenting a minimum for a particular hole-doping ($n<6$). This minimum is not at the compensated semimetal value $n=6$. In fact, its position depends on details of the model: for the tight-binding in Ref.~\cite{nosotrasprb09}, it appears for $n>6$ (not shown).  When interactions are included, $D_{ren}$ is strongly suppressed with hole doping.

The renormalization of the Drude weight can be defined as the ratio between the interacting and non-interacting values $Z_D=D_{ren}/D_{unren}$.  This quantity has been used to estimate the strength of correlations  from experiments~\cite{qazilbash09}. We discuss now why this procedure, valid for single band systems, is not directly applicable for multiorbital ones.  

At zero temperature, the optical conductivity of a system with local correlations and tight-binding $H_0
= \sum_{\bf{k}\sigma\mu\nu}\epsilon_{\mu\nu} (\bf {k})
c^\dagger_{\bf {k} \mu \sigma}c_{\bf {k} \nu \sigma} $ is~\cite{dagottoprb11,nosotrasprb13}
\begin{equation}
\sigma_\alpha^\prime(\omega)=D_\alpha\delta(\omega)
+\frac{\pi}{V}\sum_{m \neq 0} \frac{\left |\langle \phi_0 \left |j_\alpha \right |\phi_m\rangle \right|^2}{E_m-E_0} \delta(\omega-(E_m-E_0))
\label{eq:optcond}
\end{equation}
with $\alpha$ the direction and the Drude weight $D_\alpha$ 
\begin{equation}
D_\alpha=\pi \langle \phi_0|-T_\alpha|\phi_0\rangle-\frac{2\pi}{V}\sum_{m\neq0} \frac{\left|\langle \phi_0 \left |j_\alpha\right|\phi_m\rangle \right|^2}{E_m-E_0} \,,
\label{eq:drude1}
\end{equation}
with $E_0$ and $E_m$ the energies of the ground $| \phi_0 \rangle$ and excited states $|\phi_m\rangle$ and $V$ the volume. $T_\alpha=-\sum_{{\bf k}\sigma\mu\nu} \frac{\partial ^2\epsilon_{\mu\nu}(\bf {k})}{\partial k_\alpha^2}c^\dagger_{\bf {k}\mu \sigma}c_{\bf {k} \nu\sigma} $ is the diamagnetic and $j_\alpha=-\sum_{{\bf k}\sigma\mu\nu} \frac{\partial \epsilon_{\mu\nu}(\bf k)}{\partial k_\alpha}c^\dagger_{\bf {k} \mu\sigma}c_{\bf {k} \nu \sigma}$ the paramagnetic contributions.  The optical conductivity Eq.~(\ref{eq:optcond}) satisfies the restricted sum-rule~\cite{maldagueprb77,baeriswylprb87,millisprb90,dagottoreview} 
\begin{equation}
SW_{\alpha }=\int_0^\infty \sigma^\prime_{\alpha }(\omega) d\omega=\frac{\pi}{2} \langle \phi_0 \left|-T_{\alpha}\right|\phi_0\rangle={\pi \over 2} \langle -T \rangle  \, .
\label{eq:empotential}
\end{equation}
In single band systems the second terms in Eq.~(\ref{eq:optcond}) and Eq.~(\ref{eq:drude1}) account for the incoherent contribution. When hopping is restricted to first nearest neighbors $\langle T\rangle=1/2\langle K\rangle$ and the integral of the conductivity up to the interband transitions energy is proportional to the kinetic energy $K$. This fact is accounted for experimentally by integrating up to frequencies smaller than those at which the interband transitions set in~\cite{basovRMP05}.
In contrast, in multi-orbital systems the second term in Eq.~(\ref{eq:optcond}) includes both the incoherent and the interband contributions.  Therefore the integration of the low energy spectrum does not give any information on $K$.

The renormalizations of the Drude weight and of the kinetic energy have also been discussed within a Fermi liquid picture. In this case, in a single band model, and neglecting the incoherent contribution, $D=\pi\langle -T \rangle$. It is easy to see that $Z_D$ is equal to $Z_K=K_{ren}/K_{unren}$ and to the quasiparticle weight $Z$~\cite{basovreview,note-Mott} even if hopping is extended beyond nearest neighbors. This argument has been used to estimate the renormalization of the kinetic energy is several materials, including iron superconductors~\cite{qazilbash09}. However, it is not valid for a multiorbital system.  
Omitting the incoherent contribution, the Drude weight for a multiorbital system is~\cite{nosotrasprb13}
\begin{eqnarray}
D_\alpha&=&\pi  \langle -T \rangle
-\frac{2\pi}{V}\sum_{{\bf k}n \neq n'} \frac{\left |j^\alpha_{n'n}({\bf k})\right|^2}{\epsilon_{n'}({\bf k})-\epsilon_{n}
({\bf k})} \theta(\epsilon_{n'}({\bf k})) \theta(-\epsilon_{n}({\bf k})) \nonumber \\
&=& \pi \langle -T \rangle -\langle I \rangle , 
\label{eq:drude2}
\end{eqnarray}
with $\langle I\rangle$ due to interband transitions. Therefore
\begin{equation}
Z_D= \frac{\pi \langle- T_{ren}\rangle-\langle I_{ren}\rangle}{\pi\langle -T_{unren}\rangle-\langle I_{unren}\rangle} \,.
\end{equation}
If the quasiparticle weight $Z$ is equal for all orbitals the situation remains the same as in the single band case, as $Z$ can be factored out from both $\langle T \rangle$ and $\langle I \rangle$. Therefore both the Drude weight and the kinetic energy renormalizations are given by $Z$. However, in iron superconductors the quasiparticle weight is orbital dependent $Z_m$ and we cannot factor out $Z$ from these expressions.  Consequently the Drude weight and the kinetic energy are weighted by the interactions in a different way and $Z_K \neq Z_D$.

Fig.~\ref{fig:drude}(b) compares $Z_m$ with $Z_D$ and $Z_K$.  To calculate $K$ we consider 
\begin{equation}
K_{ren,unren}= \sum_{\mu\nu;{\bf i j}}\sqrt{Z_\mu Z_\nu} \,\, t_{{\bf ij}}^{\mu\nu}\langle c^{\dagger}_{\mu{\bf j}}c_{\nu{\bf i}} \rangle \,.
\nonumber
\end{equation}
$K_{ren}$ is calculated with the parameters obtained in the slave spin calculation, see Fig.~\ref{fig:drude}(b),  while $Z_m=1$ is used in the non-interacting (unrenormalized) case $K_{unren}$.

As previously discussed~\cite{yin11,liebsch2010,si2012,si2012-2,demedici2014} the most correlated orbital is $xy$ (with lobes directed towards the Fe diagonals) while the $e_g$ orbitals, $3z^2-r^2$ and $x^2-y^2$, are the least correlated ones. The suppression of $Z_m$ with hole-doping reflects the enhancement of the electronic correlations. For the explored doping range the renormalization of the Drude weight $Z_D$ fulfills $Z_{xy}<Z_D<Z_{zx/yz}$. This is consistent with the dominant presence of these three orbitals at the Fermi surface. Note that, in the hole doped case, $Z_D$ runs closer to $Z_{zx,yz}$ than to $Z_{xy}$. This is congruous with the transport being dominated by the lightest electrons (the ones in zx/yz) at the Fermi level\cite{demedici2014}.
The renormalization of the kinetic energy $Z_K$ is intermediate among all $Z_m$ and, like $Z_D$, it is also suppressed with doping. However, $Z_K > Z_D$ in all the explored doping range. It is then clear that these two quantities should not be treated as equivalent (not even in a Fermi-liquid framework).

In summary, we have shown that the interpretation of the optical conductivity spectrum of iron superconductors must take into account the orbital dependent electronic correlations, which are greatly enhanced with hole doping. We have clarified the role of interband transitions at low energies.
Although the incoherent part is expected to be present in this range of frequencies, our results emphasize that an important part of the spectral weight at low energies is due to interband transitions. The interband contribution dominates the coherent spectral weight in hole-doped samples. It is not justified to describe it in terms of a generalized Drude model or a wide (incoherent) Drude peak. 
Finally, we have also shown that both the Drude peak and the kinetic energy are strongly renormalized by electronic correlations and that both renormalizations are more pronounced with increasing hole-doping.  However the renormalization of both quantities and the information provided by them is different.  Therefore they should not be treated as equivalent. The difference between these two quantities has its origin in the orbital differentiation characteristic of iron superconductors.

We thank Ricardo Lobo and A.J. Millis for useful discussions. We acknowledge funding from Ministerio de Econom\'ia y Competitividad via Grants No.
FIS2011-29689 and FIS2012-33521. E.B. thanks the hospitality and support from ESPCI, where this work was initiated. 
\bibliography{pnictides}

\begin{thebibliography}{63}
\bibitem[{not()}]{note-Mott}
\bibinfo{note}{For a single band system the expression Eq.~(\ref{eq:drude1})
  emphasizes that the transition to a Mott insulating state with $D_\alpha$
  does not imply $\pi \langle \phi_0|-T_{\alpha}|\phi_0\rangle=0$, but $\pi
  \langle \phi_0|-T_{\alpha}|\phi_0\rangle=\frac{2\pi}{V}\sum_{m\neq0}
  \frac{\left|\langle \phi_0 \left |j_\alpha\right|\phi_m\rangle
  \right|^2}{E_m-E_0}$. Therefore, in an exact description, the kinetic energy
  does not vanish in the Mott insulator.}
\end{thebibliography}

\end{document}